# First-Principles Insights into Surface and Ligand Effects in Stoichiometric HgTe Quantum Dots

*Raagya Arora[1,*], Patrick J. Lohr[1], Dibyajyoti Ghosh[3,4], Jennifer Hollingsworth[1], Sergei Tretiak[1,2*]*

[1]*Centre for Integrated Nanotechnologies, Los Alamos National Laboratory, Los Alamos, New Mexico 87545, USA*

[2]*Theoretical Division, Los Alamos National Laboratory, Los Alamos, New Mexico 87545, USA*

[3]*Department of Chemistry, Indian Institute of Technology Delhi, Hauz Khas, New Delhi, 110016 India*

[4]*Department of Materials Science and Engineering (DMSE), Indian Institute of Technology Delhi, Hauz Khas, New Delhi, 110016 India*

*raagya@lanl.gov, serg@lanl.gov

## Abstract

HgTe quantum dots are promising mid-infrared nanomaterials owing to their exceptional bandgap tunability, yet their electronic structure is strongly influenced by surface coordination and ligand passivation at ultrasmall sizes. Here, we employ atomistic simulations to systematically investigate stoichiometric HgTe nanoclusters with sizes ~0.86–1.85 nm. The *in-silico* exploration uncovers a transition from confinement-dominated electronic structures with delocalized frontier states in small self-passivated clusters to surface-influenced characteristics in larger nanoclusters. Increased coordination and bond-length inhomogeneity in the larger nanoclusters generate localized near-gap states centered on undercoordinated surface atoms. At intermediate sizes, the band edge states become spatially separated on different regions of the cluster without forming deep gap states, marking the onset of surface-induced electronic asymmetry. In larger clusters (~1.8 nm), common neutral ligands like amines, thiols, phosphines, and alcohols effectively eliminate surface-derived localized states by restoring local coordination and altering the band-edge electronic structure through ligand-surface hybridization. The sensitivity of the bandgap to ligand identity and binding site underscores the interplay between surface coordination and ligand chemistry in shaping the electronic structure of these nanoclusters. These insights provide an atomistic understanding of size-dependent electronic structures in ultrasmall HgTe clusters. The study further establishes neutral ligands as powerful chemical handles for engineering frontier electronic states relevant to infrared optoelectronic functionality.

## Introduction

Colloidal quantum dots (QDs) have emerged as versatile building blocks for optoelectronics due to the exceptional tunability of their electronic and optical properties via control over size, composition, shape, and surface chemistry.[1–3]Such remarkable flexibility enables a wide range of applications, including light emission, photodetection, photovoltaics, and infrared sensing.[2,4,5] However, as nanocrystals approach ultrasmall dimensions, their properties are no longer governed by quantum confinement alone. Due to the large fraction of atoms residing at the surface, the frontier electronic structure of these ultrasmall QDs can be strongly influenced by local coordination, surface reconstruction, ligand chemistry, and their dynamics.[6–9] Therefore, understanding the complex interplay between confinement and surface chemistry is essential for precisely controlling the optoelectronic properties of QDs in the ultrasmall regime.

HgTe QDs are particularly compelling as their bulk form is a semimetal with inverted band ordering, whereas quantum confinement opens a finite gap at the nanoscale.[10–14] This duality allows HgTe quantum dots to access the near-, mid-, and long-wave infrared spectral ranges, making those promising materials for infrared photodetectors, emitters, and imaging technologies.[10–13] Experimentally, colloidal HgTe QDs have demonstrated size-tunable absorption and emission from the near-infrared into the mid-infrared, with photodetection demonstrated beyond 5 μm and extending to 12 μm in larger nanocrystals.[11–13] Recent reviews further emphasize that HgTe occupies a unique position among colloidal QD materials due to this exceptional spectral tunability, ranging from the visible/near-infrared toward the terahertz regime.[5,15,16]

An increasing attention has recently been directed toward ultrasmall HgTe quantum dots (<2-3 nm), which occupy the extreme confinement regime and exhibit optical gaps in the near- and short-wave infrared (NIR/SWIR) spectral range.[5,10,17] Recent experimental studies have demonstrated that such ultrasmall HgTe nanocrystals can retain high photoluminescence quantum yields while providing access to technologically important NIR/SWIR wavelengths relevant to imaging, sensing, and optoelectronic applications.[17] Despite this growing interest, considerably less is known about the atomistic electronic structure of ultrasmall HgTe clusters than about their larger mid-infrared counterparts.[12–14,18] In this size regime, where a large fraction of atoms reside at or near the surface, and the distinction between core and surface begins to break down, electronic properties are expected to be governed not only by quantum confinement but also by local coordination, surface reconstruction, and ligand chemistry.[3,6,8,9,19–21]

The theoretical modeling provides initial details of the electronic and optical properties of HgTe QDs. Tight-binding method-based calculations describe the role of quantum confinement in modifying the inverted electronic structure of HgTe nanocrystals and assign size-dependent optical transitions.[22,23] Experimental spectroelectrochemical studies further reveal conduction-band fine structure in HgTe QDs, with spin-orbit coupling and reduced symmetry contributing to the splitting of intraband transitions.[18] Despite providing valuable insights into the optical spectra of HgTe QDs, these studies do not fully resolve the impact of local surface coordination in governing the localization of frontier electronic states in ultrasmall stoichiometric clusters.

The detailed surface chemistry plays a crucial role in determining the optoelectronics of semiconductor QDs. In its colloidal form, ligands provide solubility and colloidal stability to QDs, but they also passivate undercoordinated atoms, regulate surface stoichiometry, modify interfacial dipoles, and hybridize with band-edge states.[3,24–28] Prior theoretical work on II-VI semiconductor nanocrystals has shown that surface coordination plays a decisive role in the formation of localized near-gap states, and that ligand binding or ligand loss can strongly alter the electronic structure near the highest occupied molecular orbital (HOMO)- lowest unoccupied molecular orbital (LUMO) gap.[25–28] These studies establish that the surface is not a passive boundary of a confined semiconductor core, but an electronically active interface.

Surface effects are expected to be especially pronounced in HgTe as the low-energy electronic structure is narrow-gap, relativistic, and sensitive to local bonding. Experimental studies have shown that HgTe QD surface composition can influence doping behavior, with air-stable n-type character in large HgTe QDs attributed to Hg-rich surfaces.[23] Recent first-principles work on HgTe(111) surface models has also identified unpassivated surface Hg atoms as a source of trap states and showed that chemical passivation can mitigate these states.[24] However, idealized surface models do not capture the full structural heterogeneity of finite ultrasmall clusters, where the surface consists of multiple inequivalent local environments rather than extended periodic facets. Moreover, relatively little attention has been devoted to HgTe clusters in this ultrasmall size regime, despite the expectation that surface coordination, reconstruction, and quantum confinement effects should be particularly pronounced. These considerations motivate a cluster-based approach that explicitly accounts for the diversity of local surface environments and their influence on electronic structure.

Several questions, therefore, remain open for stoichiometric HgTe QDs in the ultrasmall regime. Does quantum confinement alone determine the frontier electronic structure, or does surface coordination become dominant even in neutral stoichiometric clusters? At what size does the electronic structure evolve from delocalized, core-like frontier states to surface-influenced localized edge states? Can such states arise from coordination and bond-length heterogeneity without invoking non-stoichiometry, charge injection, or ligand loss? How do neutral ligands tune these states through local coordination, ligand-surface hybridization, and adsorption-site dependence?

Here, we address these questions using density functional theory (DFT)-based calculations on stoichiometric HgTe nanoclusters containing 14, 20, 38, 82, and 86 atoms, corresponding to diameters of approximately 0.86-1.85 nm. These clusters were selected as representative stoichiometric structures spanning the ultrasmall confinement regime, including sizes comparable to recently reported ultrasmall HgTe quantum dots containing on the order of 100 atoms.[17] We analyze structural relaxation, density of states, projected density of states, real-space frontier orbitals, and inverse participation ratio to connect local coordination environments with electronic localization. The calculations reveal three regimes: ultrasmall clusters with clean confinement-induced HOMO-LUMO gaps and delocalized frontier states; an intermediate regime in which the HOMO and LUMO become spatially polarized without forming deep gap states; and larger clusters in which surface coordination and bond-length heterogeneity produce localized, surface-associated edge states. We then examine neutral ligand passivation of the 86-atom cluster using methylamine, methanethiol, methylphosphine, and methanol as representative donor ligands. These atomistic simulations show that ligand passivation suppresses surface-localized edge states while also tuning frontier levels in a ligand- and site-dependent manner. Additional spin-orbit coupling and hybrid-functional calculations for the 86-atom cluster assess the robustness of the surface-influenced frontier electronic structure beyond scalar-relativistic semilocal DFT. These detailed insights establish an atomistic picture of quantum confinement, surface coordination, and neutral ligand chemistry together controlling the electronic structure of ultrasmall stoichiometric HgTe QDs.

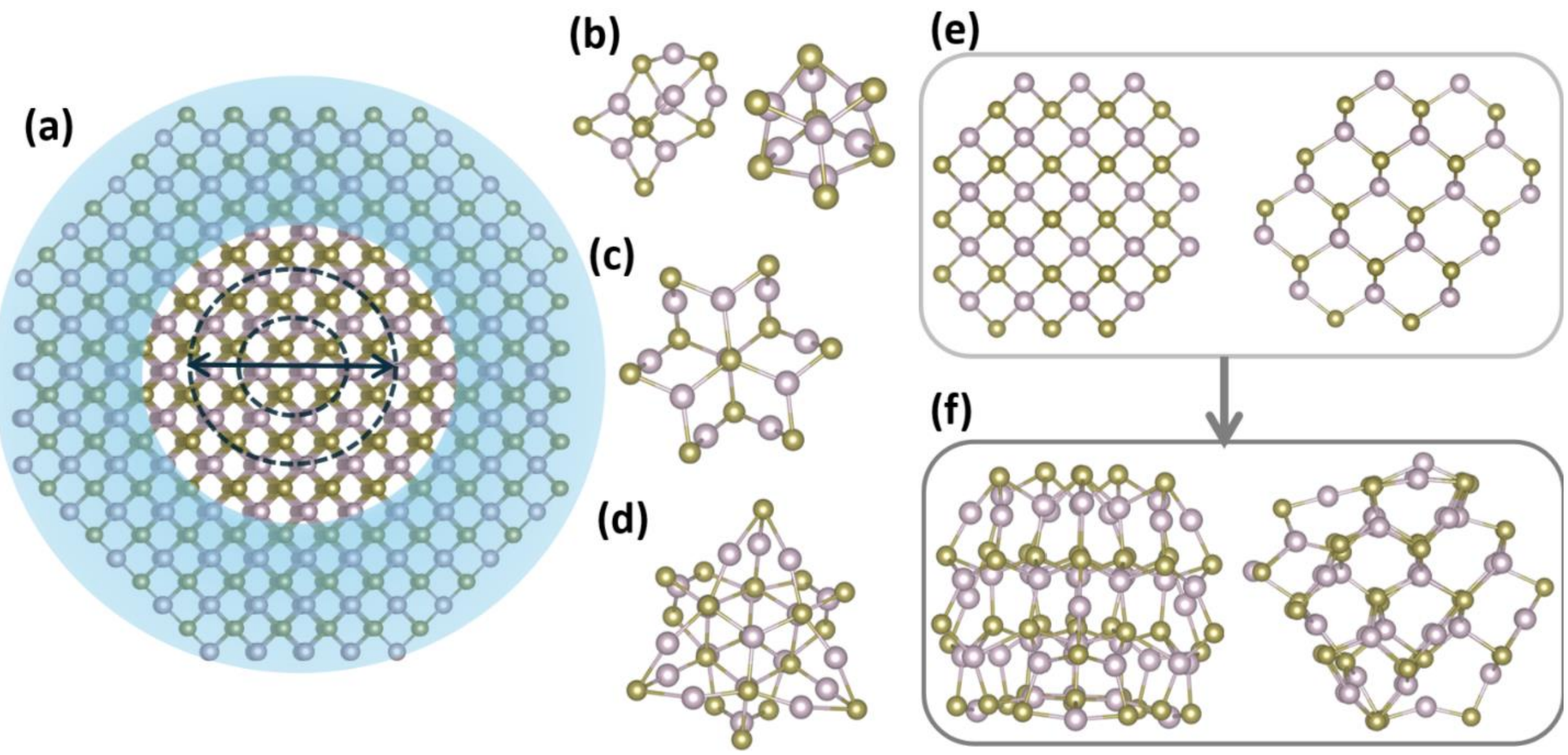

**Figure 1:** (a) Schematic representation of the cluster-generation procedure, where stoichiometric nanoclusters are obtained by spherical truncation of bulk zincblende HgTe. (b–d) Relaxed geometries of the 14-, 20-, and 38-atom HgTe clusters, corresponding to ultrasmall quantum dots in the confinement-dominated regime. (e) Initial truncated geometry and (f) DFT-relaxed structure of the 86-atom HgTe cluster, showing relaxation-induced surface reconstruction and increased structural heterogeneity at the outer shell.

## Results and Discussion

To examine how confinement and surface coordination jointly control the electronic structure of ultrasmall HgTe quantum dots, we construct a series of stoichiometric HgTe nanoclusters from bulk zincblende HgTe. A radius-based truncation scheme generates approximately spherical clusters (Figure 1a). We explore several centering choices, including geometric-center, Hg-centered, Te-centered, and Hg-Te bond-midpoint constructions to sample different surface terminations and local coordination environments. We enforce the stoichiometry by enforcing equal numbers of Hg and Te atoms in selected clusters. The final size series contains 14, 20, 38, 82, and 86 atoms, corresponding to approximate diameters of 0.86, 1.10, 1.28, 1.72, and 1.85 nm, respectively (Figure 1b-f). This size range allowed us to follow the evolution from strongly confined molecular-like clusters to larger nanoclusters in which surface coordination begins to dominate the frontier electronic structure. (Figure 1)

Before analyzing the electronic states, we first assessed whether the relaxed clusters preserve a chemically meaningful Hg-Te bonding framework. Across the size series, structural relaxation maintained a predominantly heteropolar Hg-Te network rather than producing metallic or homopolar reconstruction (Figure S1, S2, S3, S4). In the smaller 14- and 20-atom clusters, nearest-neighbor Hg–Te bond lengths remained concentrated within approximately 2.6-3.0 Å, consistent with zincblende-like HgTe bonding (Figure S1, S2). No anomalously short Hg-Te contacts below ~2.5 Å or strongly elongated bonds above ~3.5 Å were observed, indicating that relaxation did not induce unphysical bond collapse or bond breaking. Importantly, same-species separations remained well outside bonding range (Te-Te > ~4.5 Å; Hg-Hg > ~3.1 Å), confirming that the clusters retained a predominantly heteropolar bonding topology.

The primary effect of structural relaxation is therefore not bond breaking, but the redistribution of local coordination environments within the nanoclusters. Hg atoms preferentially occupy more internal positions, while Te atoms enrich the outer shell, producing an effectively Te-rich surface. This radial redistribution reduced the number of exposed undercoordinated Hg sites and provided a structural mechanism for self-passivation in the smallest clusters (Figure S1, S2). As the cluster size increased, however, the interior became more coherent while the outer shell developed a broader distribution of coordination numbers and Hg-Te bond lengths. Thus, we observed that relaxation played a dual role: it stabilized the smallest clusters by suppressing highly reactive dangling-bond-like motifs, but it also generated increasing surface heterogeneity in the larger clusters. We do not claim that Te termination is the thermodynamic equilibrium surface of HgTe. Rather, we observe a consistent relaxation-induced radial segregation in the stoichiometric clusters studied here, which correlates with suppression of undercoordinated Hg sites and the emergence of self-passivated electronic structure.

This structural evolution motivated the use of a localization metric to distinguish delocalized confinement-derived states from surface-associated frontier states. We quantified localization using the inverse participation ratio,

$$\mathrm{IPR} = \sum \rho_i^2 \Big/ \left(\sum \rho_i\right)^2 \tag{1}$$

where $\rho_i$ is the atom-projected contribution of atom $i$ to a given Kohn-Sham state. For a state uniformly delocalized over $N$ atoms, IPR $\approx 1/N$, whereas larger values indicate increasing localization on a smaller subset of atoms. In the analysis below, low IPR values-along with spatially distributed wavefunctions-were used to identify delocalized frontier states, while elevated IPR values and real-space localization on specific surface regions indicated surface-associated edge states.

Having established that the relaxed clusters retained a stable heteropolar framework while developing size-dependent surface heterogeneity, we next examined how this structural evolution is reflected in the density of states, orbital character, and frontier-state localization.

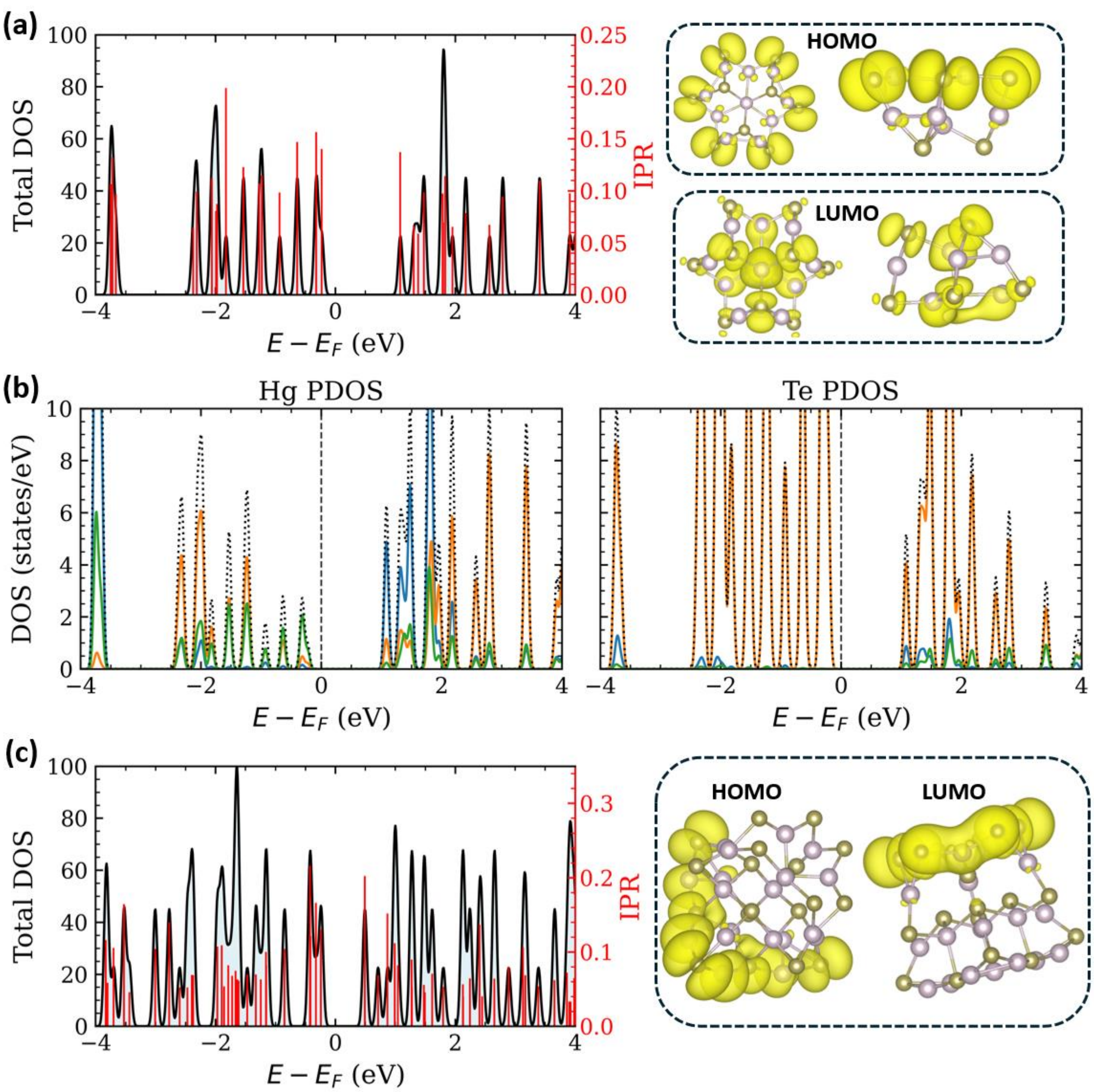

**Figure 2:** (a) Total density of states (DOS, black) and inverse participation ratio (IPR, red) for the 20-atom HgTe cluster, together with real-space HOMO and LUMO isosurfaces. The 20-atom cluster exhibits a clean frontier gap and relatively delocalized HOMO/LUMO distributions, indicating confinement-dominated electronic structure without strongly localized near-gap states. (b) Atom- and orbital-projected DOS for the 20-atom cluster, resolved into Hg and Te contributions and separated into s, p, and d components. The occupied frontier states are dominated primarily by Te 5p-character, while the unoccupied states show stronger Hg-derived contributions with additional Te 5p-hybridization. (c) Total DOS, IPR, and HOMO/LUMO isosurfaces for the 38-atom HgTe cluster. Although the cluster retains a clean frontier gap, the HOMO and LUMO become spatially polarized on different regions of the nanocluster, accompanied by enhanced IPR values near the band edges. This indicates the onset of surface/electrostatic asymmetry and partial frontier-state localization before the formation of strongly surface-localized edge states in larger clusters.

### Quantum Confinement in Small HgTe Clusters

The electronic structures of the representative small and intermediate clusters are summarized in Figure 2. For the 14- (Figure S5, S6) and 20-atom clusters (Figure 2a, b), the calculated density of states consisted of discrete, well-separated levels with a clean HOMO-LUMO gap. This molecular-like electronic structure reflects strong quantum confinement and the suppression of bulk-like semimetallic behavior. Importantly, despite the large fraction of surface atoms, no mid-gap states were observed in these smallest stoichiometric clusters.

The absence of mid-gap states indicated that structural relaxation effectively suppresses electronically active undercoordinated motifs. In this regime, the HOMO and LUMO are spatially distributed across the cluster rather than localized on isolated surface sites, and the corresponding IPR values remain relatively low (Figure 2a). The resulting HOMO-LUMO gap is therefore best described as arising from quantum confinement within a self-passivated heteropolar framework, rather than from confinement alone or from isolated dangling-bond-like surface states. Projected density of states (pDOS) analysis further supports the interpretation. We found the occupied frontier states to be primarily Te-derived, with dominant Te p-character, whereas the unoccupied states contained stronger Hg-derived contributions with additional Hg-Te hybridization (Figure 2b). This chemically interpretable orbital character indicates that the frontier states originate from the confined HgTe framework rather than from localized surface defects. Thus, in the ultrasmall limit, stoichiometric HgTe clusters retain clean frontier gaps and delocalized cluster-derived frontier states even though nearly all atoms are close to the surface.

### Intermediate Regime (38-atom cluster)

The 38-atom cluster represents an intermediate regime between fully delocalized and surface-influenced electronic structure. While a clean HOMO-LUMO gap was retained upon structural relaxation, the frontier orbitals exhibited spatial asymmetry, with the HOMO and LUMO localized on opposite regions of the cluster (Fig. 2c). This behavior arises from structural asymmetry and the heteropolar nature of Hg-Te bonding. Valence states preferentially localized on Te-rich regions, while conduction states are localized on Hg-rich regions. In the absence of bulk symmetry, this leads to an internal electrostatic potential gradient, effectively producing a dipole-like polarization across the cluster. The 38-atom cluster represents an intermediate regime, as shown by its DOS/IPR profile and polarized HOMO/LUMO isosurfaces in Figure 2c. Importantly, this spatial separation does not correspond to defect or surface-associated edge states, as the gap remains clean. Instead, it reflects intrinsic charge separation driven by finite-size effects. This regime marks the onset of surface influence on electronic structure, preceding the formation of surface-localized states in larger clusters. This indicates that structural and electrostatic asymmetry can perturb the band edges before true near-gap surface states appear. Thus, the onset of localization in HgTe does not initially occur through deep trap formation, but through polarization of otherwise gap-preserving frontier states.

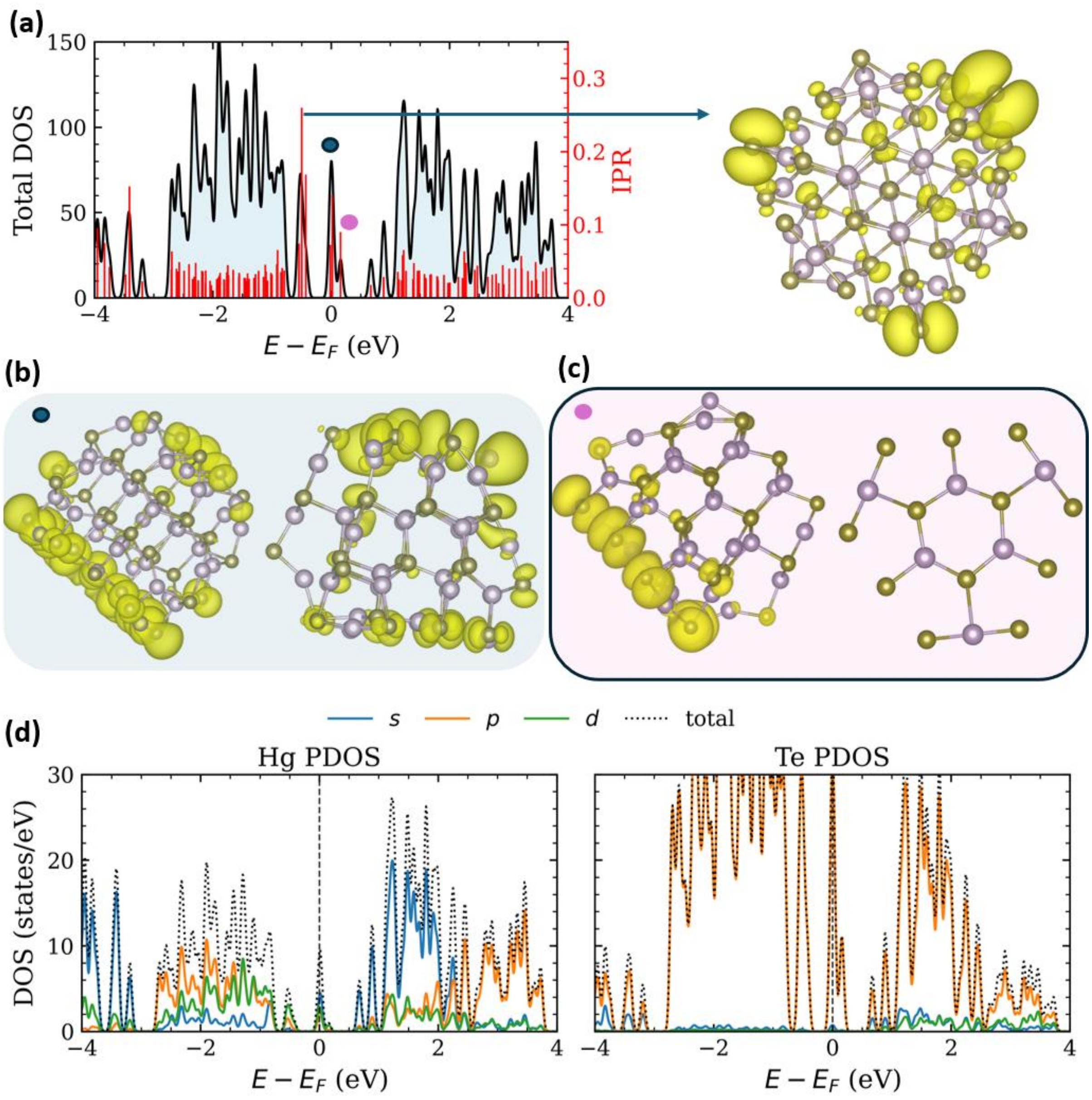

**Figure 3.** (a) Total density of states (DOS, black) and inverse participation ratio (IPR, red) for the relaxed 86-atom HgTe cluster. In contrast to the smaller clusters, the 86-atom cluster exhibits discrete states close to the frontier region together with enhanced IPR values, indicating increased localization of near-edge electronic states. The colored markers identify representative states whose real-space wavefunctions are shown in panels (b) and (c). (b, c) Real-space isosurfaces of selected near-frontier states corresponding to the marked DOS/IPR features in panel a. The wavefunctions are localized on distinct surface regions rather than distributed uniformly throughout the cluster, revealing the role of facet-like surface domains and local coordination heterogeneity in generating surface-associated edge states. (d) Atom- and orbital-projected density of states (PDOS) for Hg and Te atoms.

**Emergence of Surface States in Larger Clusters**

To distinguish core and surface bonding environments in the 86-atom HgTe cluster, atoms were classified according to both radial position and Hg-Te coordination number. This analysis identifies a compact interior of near-tetrahedral sites surrounded by a large population of low-coordination surface atoms. The mean radial distance is larger for Te than for Hg, indicating that Te remains preferentially enriched at the outer shell. Core-core Hg-Te bonds are comparatively narrow in distribution (2.84-2.96 Å), whereas surface-involving bonds span a much broader range, particularly for subsurface-surface contacts (2.685-3.241Å). Thus, the dominant structural signature of the 86-atom cluster is not a simple surface bond contraction, but rather substantial bond-length inhomogeneity at the outer shell. These structural changes are reflected directly in the electronic structure of the 86-atom cluster (Figure 3).

As the cluster size approaches ~1.8 nm (82-86 atoms), additional electronic states emerged near the band edges due to the increasing population of undercoordinated surface atoms and the heterogeneous bonding environment at the interface. As shown in Figure 3a, the 86-atom cluster exhibits discrete near-frontier states accompanied by elevated IPR values. In contrast to the smaller clusters, where frontier states remain largely delocalized and core-derived, the band-edge orbitals of the larger clusters acquire substantial surface character, as indicated by elevated IPR values and real-space wavefunction localization.

The broader distribution of surface Hg-Te bond lengths perturbs the energies of surface-centered orbitals, producing shallow localized edge states associated with incomplete coordination and bond-length inhomogeneity. These states are spatially confined to specific surface regions rather than delocalized throughout the cluster interior, indicating that the surface increasingly contributes to the low-energy electronic structure. The emergence of these surface-associated edge states is expected to reduce electron-hole overlap relative to the smaller clusters and increase the sensitivity of the electronic structure to local surface chemistry. The transition from confinement-dominated to surface-influenced electronic structure in HgTe clusters is therefore governed primarily by increasing structural and coordination heterogeneity at the surface. The corresponding pDOS indicates substantial Te p- and Hg-s derived contributions near the frontier region (Figure 3d).

As HgTe contains heavy elements, SOC calculations were performed for the 86-atom cluster to assess the robustness of the frontier electronic structure under relativistic effects. These calculations show that the surface-influenced frontier electronic structure is not eliminated by relativistic effects (Figure S9). SOC modifies the distribution of states near the frontier region and produces a partially depleted low-DOS region around the Fermi level rather than a fully clean gap. pDOS analysis shows that states near the occupied edge remain predominantly Te-derived, while Hg contributions remain significant near the unoccupied edge, consistent with the expected orbital character of confined HgTe. These results indicate that SOC perturbs the near-edge energetic structure but does not eliminate the surface-influenced frontier-state character observed at the scalar-relativistic PBE-based calculations.

Hybrid-functional calculations were also performed for the bare 86-atom cluster as a higher-level check on the PBE description. These calculations consistently exhibit a less crowded but still surface-influenced frontier region (Figure S10). The hybrid-functional DOS shows a more discrete and less crowded frontier region, indicating that semilocal DFT likely overestimates the density of near-edge states. However, the persistence of surface-influenced frontier features supports the conclusion that the large-cluster electronic structure is governed by surface coordination and bond-length heterogeneity rather than by semilocal functional error alone.

Real-space wavefunction visualizations confirm that these states are localized on specific surface regions rather than delocalized throughout the cluster interior (Figure 3b, c). The corresponding frontier orbitals are not uniformly distributed over the surface but are instead localized on specific regions of the nanocluster, corresponding to distinct facet-like domains and coordination environments. These regions are characterized by lower coordination numbers and increased bond-length distortion relative to the interior.

The spatial localization of near-gap states can be further interpreted in terms of facet-like surface regions derived from the underlying zincblende structure. Although the nanoclusters do not exhibit perfectly defined crystallographic facet-like domains due to structural relaxation, the surface can still be described in terms of locally ordered regions resembling low-index planes. In particular, the observed localization of frontier orbitals is consistent with facet-like regions resembling polar (111)-like motifs, where Hg-terminated and Te-terminated domains are exposed. These polar surface motifs naturally give rise to differences in local electrostatic potential and coordination environment, leading to preferential localization. The presence of such facet-dependent polarity provides an additional driving force for the formation of surface-localized electronic states in larger clusters.

**Ligand Passivation of Surface States:**

Ligand passivation of the 86-atom cluster is summarized in Figure 4. The effect of ligand passivation is strongly dependent on the local surface environment of the nanocluster. In the unpassivated 86-atom HgTe cluster, wavefunction visualizations reveal that near-gap states are not uniformly distributed across the surface but are localized on specific regions. Based on this spatial localization, two representative surface regions, denoted as Face 1 and Face 2, are identified as electronically active sites where undercoordinated atoms contribute most strongly to the band-edge electronic structure (Figure 4). It is important to note that these regions do not represent a complete set of crystallographic facets. Due to the small cluster size and structural relaxation, the surface is composed of multiple inequivalent local environments rather than extended periodic planes. However, the selected adsorption regions retain local structural motifs that are analogous to those found on zinc-blende HgTe (111) surfaces, including both Hg-rich and Te-rich coordination environments. Face 1 and Face 2, therefore, serve as representative models of electronically distinct surface domains that may occur on the nanocluster, rather than as unique or exhaustive surface types.

The representative methylamine-passivated structures and corresponding DOS/IPR plots show that ligand coordination suppresses localized near-frontier states in a site-dependent manner (Figure 4a). To quantify the thermodynamic stabilization associated with ligand passivation, we calculated the relaxed passivation energy,

$$E_{pass}(n,L)=E(QD+nL)-E(QD)-nE(L),$$

where $E(QD+nL)$ is the total energy of the relaxed ligand-passivated cluster, $E(QD)$ is the energy of the relaxed bare cluster, $E(L)$ is the energy of the isolated ligand, and $n$ is the number of ligands bound to the cluster. More negative values indicate greater energetic stabilization relative to the separated bare cluster and free ligands.

The calculated binding energies are negative for all ligand-passivated structures, indicating that ligand coordination is energetically favorable and stabilizes the 86-atom nanocluster. For the fully passivated six-ligand configurations, the binding energies vary only modestly among the ligand types, ranging from approximately −1.56 to −1.61 eV per ligand. Methylamine exhibits the strongest binding, followed by methylphosphine, methanethiol, and methanol, suggesting that all four ligands provide comparable thermodynamic stabilization of the cluster surface. In contrast, substantially larger variations are observed as a function of ligand coverage and adsorption site. For both methylamine and methanethiol, the binding energy becomes increasingly negative as the number of adsorbed ligands decreases, indicating that ligand–ligand interactions and steric crowding weaken adsorption at high surface coverage. Despite these trends, the binding energies do not directly correlate with the observed HOMO-LUMO gaps. For example, ligands with similar adsorption strengths can produce substantially different electronic gaps, indicating that frontier-level shifts are governed not only by adsorption energetics but also by ligand orbital alignment, ligand-surface hybridization, and local electrostatic effects.

Upon ligand binding, these electronically active regions become coordinated, stabilizing coordination-deficient motifs associated with the near-edge states and recovering a cleaner frontier gap region. Such insights demonstrate that the origin of these states is local, arising from specific coordination-deficient surface environments. Structurally, ligand adsorption reduces bond-length dispersion at the surface and stabilizes local coordination, thereby decreasing the heterogeneity of the outer shell. Electronically, this suppresses surface-localized orbitals by redistributing charge density away from undercoordinated sites, effectively removing or shifting near-gap states away from the band edges. Similar DOS trends are observed for methanethiol, methylphosphine, and methanol passivation (Figure 4b-d). Thus, ligand passivation operates through two coupled mechanisms. First, it restores local coordination by saturating undercoordinated surface atoms, suppressing surface-localized edge states. Second, it modifies the interfacial electronic structure through ligand-surface orbital hybridization and local electrostatic effects. As a result, the band edges are not only cleared of localized surface-derived states but also shifted in a ligand-dependent manner.

Ligand binding to different surface regions produces distinct electronic responses. For a given ligand, adsorption on Face 1 and Face 2 can yield different HOMO-LUMO gaps, reflecting variations in local coordination environment, bonding geometry, and surface electrostatics. These differences indicate that the electronic impact of passivation depends not only on ligand chemistry but also on the specific surface site to which the ligand binds. Surface regions associated with localized frontier states in the bare cluster are particularly sensitive to ligand-induced perturbations, as passivation modifies or removes the corresponding surface-derived electronic states. The observed face-dependent variations in both binding energy and HOMO-LUMO gap therefore highlight the importance of local surface heterogeneity in determining the electronic properties of ultrasmall HgTe clusters.

Partial ligand coverage further highlights such heterogeneity. Incompletely passivated clusters retain undercoordinated sites distributed across multiple surface regions, leading to residual near-gap states and non-monotonic bandgap trends. The resulting HOMO-LUMO gaps depend on both ligand identity and adsorption site (Figure 4f). These results demonstrate that ligand effects in ultrasmall quantum dots are intrinsically site-dependent and cannot be fully described in terms of idealized facets. Instead, the electronic response is governed by a distribution of local coordination environments, with ligand passivation acting selectively on the most electronically active regions.

**Effect of Ligand Chemistry**

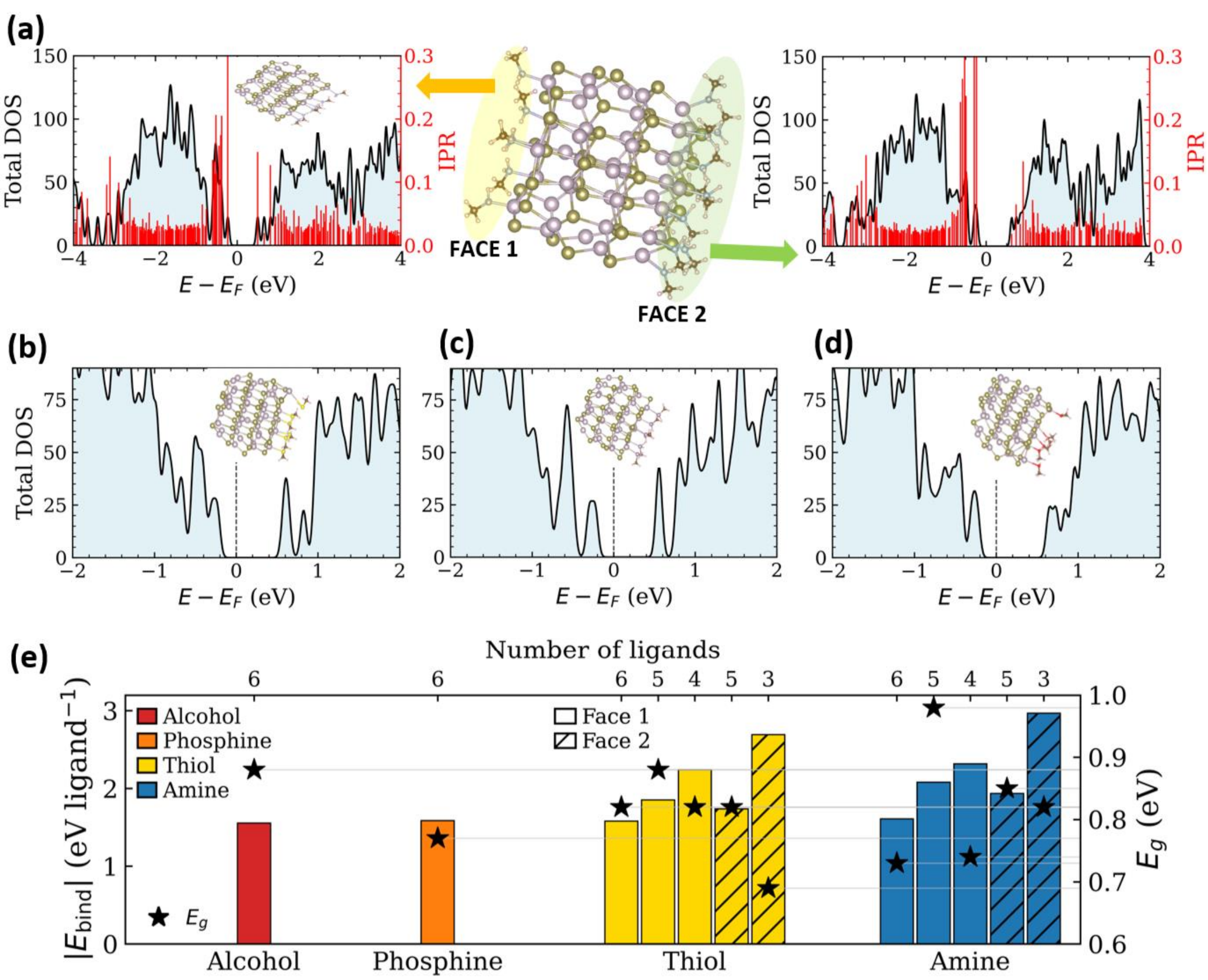

**Figure 4.** (a) Methylamine-passivated 86-atom HgTe cluster with ligands coordinated to undercoordinated surface Hg sites on two representative surface regions, together with the corresponding DOS/IPR plots. Black curves denote total DOS and red bars denote IPR; (b-d) Total DOS for methanethiol-, methylphosphine-, and methanol-passivated clusters, respectively, with relaxed structures shown as insets. Ligand coordination suppresses surface-associated near-frontier states and produces ligand-dependent frontier gaps. (e) Calculated binding energies and HOMO-LUMO gaps as a function of ligand chemistry, ligand coverage, and adsorption site. Bar heights represent the magnitude of the binding energy per ligand, with solid bars corresponding to Face 1 adsorption geometries and hatched bars corresponding to Face 2 adsorption geometries. Bar colors distinguish ligand chemistry (alcohol, phosphine, thiol, and amine), while black stars indicate the corresponding HOMO–LUMO gaps. Ligand counts are shown above each bar.

The calculated band gaps for the Face 1 ligand-passivated 86-atom HgTe cluster follow the trend methanol (0.88 eV) > methanethiol (0.82 eV) > methylphosphine (0.77 eV) > methylamine (0.73 eV). All four ligands effectively passivate undercoordinated surface atoms and eliminate surface-derived near-gap states, resulting in a clean band gap in each case. The remaining differences in band gap arise from variations in ligand-induced perturbations of the HgTe frontier electronic structure. As shown by the ligand-projected density of states (Figure S46), oxygen-derived states are located deepest within the valence region and exhibit minimal spectral weight near the band edges, indicating relatively weak coupling to the frontier HgTe states. Consequently, methanol preserves the largest band gap. In contrast, nitrogen-derived states occur at higher energies and extend closest to the valence-band edge, consistent with stronger ligand-surface hybridization and a greater perturbation of the frontier electronic structure, resulting in the smallest band gap. Sulfur- and phosphorus-derived states occupy intermediate energy ranges and produce correspondingly intermediate band gaps. Overall, the observed trend suggests that band-gap reduction correlates with the energetic proximity of ligand-localized states to the HgTe band-edge manifold and the associated degree of ligand-surface hybridization.

The ligand-dependent bandgap trends are expected to correlate with ligand-surface binding strength. Stronger binding generally enhances orbital hybridization and charge redistribution at the interface, leading to larger perturbations of band-edge states. However, binding energy alone does not fully determine the electronic response, as orbital character and energetic alignment also play critical roles. Thus, bandgap tuning reflects a combined effect of coordination strength and ligand-surface electronic coupling.

The size-dependent evolution of the frontier electronic structure is expected to have important implications for the optical behavior of ultrasmall HgTe nanoclusters.[12,13]Delocalized frontier states in the smallest clusters are associated with strong electron-hole overlap and confinement-dominated optical behavior. In contrast, the spatial separation of occupied and unoccupied frontier states in the 38-atom cluster indicates reduced overlap, even in the absence of defect-like mid-gap states. In the larger clusters, surface-associated edge states introduce localized frontier character and increase the sensitivity of the electronic structure to local surface chemistry. Ligand passivation suppresses this surface-induced electronic inhomogeneity and can therefore modify both the energetic spacing and spatial character of frontier transitions.[8,9,19,20]

## Conclusions

In this work, we have investigated the size- and ligand-dependent electronic structure of stoichiometric HgTe nanoclusters using first-principles calculations. The results reveal a clear transition from confinement-dominated electronic structure in the smallest clusters to surface-influenced frontier-state localization in larger clusters. In the 14-20 atom clusters, structural relaxation preserves a heteropolar Hg-Te framework and produces clean wide-open HOMO-LUMO gaps with delocalized frontier states, consistent with effective self-passivation. The mid-sized 38-atom cluster retains a clean gap but develops spatially polarized frontier orbitals, marking the onset of surface-induced electronic asymmetry. In the larger 82-86 atom clusters, coordination and bond-length heterogeneity at the surface introduce shallow surface-associated edge states localized on specific surface regions. SOC and hybrid-functional calculations for the 86-atom cluster indicate that this surface-influenced frontier electronic structure is not eliminated by relativistic effects or hybrid exchange, although the detailed energetic distribution of near-edge states is method-dependent. Common neutral ligands passivate and suppresses surface-localized edge states by stabilizing undercoordinated surface atoms and reducing local electronic inhomogeneity. At the same time, ligand binding shifts the frontier levels through ligand-surface hybridization and local electrostatic effects, producing ligand- and site-dependent HOMO-LUMO gaps.

Overall, these results show that the electronic structure of ultrasmall HgTe quantum dots is governed by the coupled interplay of quantum confinement, surface coordination, and ligand chemistry. We show that ligands act not merely as passive stabilizing agents, but as active chemical handles for controlling band-edge energies, wavefunction localization, and near-gap electronic structure. These insights are relevant to the design of HgTe-based infrared photodetectors, emitters, and quantum optoelectronic devices, where surface-associated states critically influence carrier trapping, recombination, and transport.

## Computational Methods

### DFT Methodology

All density functional theory (DFT) calculations were performed using the Vienna *Ab initio* Simulation Package (VASP).[29,30] The projector-augmented wave (PAW) method[31,32] was employed together with the Perdew-Burke-Ernzerhof (PBE) generalized gradient approximation exchange-correlation functional.[33] A plane-wave cutoff energy of 450 eV was used throughout. Long-range dispersion interactions were accounted for using Grimme's DFT-D3 semiempirical van der Waals correction with Becke-Johnson damping (D3(BJ)).[34,35]

Self-consistent electronic iterations were converged to $10^{-7}$ eV to ensure accurate determination of frontier energy levels and density of states. Structural relaxations were performed using the conjugate-gradient algorithm until residual atomic forces were below 0.02 eV $Å^{-1}$. Electronic occupations were treated using Gaussian smearing with a width of 0.05 eV. Additional calculations with different smearing widths confirmed that the qualitative electronic structure trends were insensitive to the smearing parameter.

A vacuum region of at least 20 Å was included in all Cartesian directions to eliminate interactions between periodic images. Dipole corrections were applied in all directions to account for asymmetry in the charge distribution of the finite clusters.[36] Electronic structure analysis was carried out using total and projected density of states (DOS/PDOS), real-space wavefunction visualization, and inverse participation ratio (IPR) calculations. The DOS was used to identify the HOMO-LUMO gap and the distribution of near-frontier states, while PDOS analysis resolved the orbital contributions of Hg and Te atoms. Real-space visualization of frontier orbitals was used to distinguish delocalized cluster-derived states from localized surface-associated edge states.

Additional calculations including spin-orbit coupling (SOC) were performed for the 86-atom HgTe cluster in order to assess the influence of relativistic effects on the frontier electronic structure. Because Hg and Te are heavy elements with strong relativistic character, SOC was included self-consistently within the VASP framework using the PBE functional.[37,38] Total and projected DOS were analyzed to evaluate the effect of SOC on the near-frontier electronic structure and orbital character.

### Gaussian Calculations

Additional electronic structure calculations were performed using the Gaussian electronic structure package[39] to evaluate basis-set and solvation effects on the electronic structure of HgTe nanoclusters. The PBE exchange-correlation functional[33] was employed together with the LANL2DZ basis set and effective core potential (ECP),[40] as well as the def2-SVP split-valence basis set with polarization functions.[41] The use of ECPs enables efficient inclusion of scalar relativistic effects, which are important for accurately describing heavy elements such as Hg and Te. All Gaussian calculations were carried out on geometries optimized using VASP, ensuring consistency between the plane-wave and localized-basis descriptions.

Solvent effects were incorporated using implicit continuum solvation models. In particular, the integral-equation-formalism polarizable continuum model (IEF-PCM)[42–44] with water as the dielectric medium was used to capture electrostatic stabilization of ligand-passivated clusters. Standard PCM calculations were also performed for comparison, while the SMD solvation model[45] was employed to include both electrostatic and non-electrostatic contributions such as cavitation and dispersion.

HOMO–LUMO gaps, orbital character, and frontier-level shifts were analyzed across gas-phase and solvated environments to assess the influence of basis set and dielectric screening on the electronic structure of HgTe nanoclusters.

Additional hybrid-functional calculations were performed for the bare 86-atom stoichiometric HgTe cluster using the screened Heyd-Scuseria-Ernzerhof hybrid functional as implemented in Gaussian (HSEH1PBE), which incorporates 25% short-range Hartree-Fock exchange.[46,47] The 86-atom cluster was selected because it exhibits the strongest surface influence on the frontier electronic structure among the stoichiometric clusters considered here. Single-point hybrid-functional calculations were performed on the PBE-relaxed geometry to assess whether the shallow near-frontier states identified at the semilocal level remain robust upon inclusion of nonlocal exchange.

**Ligand Passivation**

The effect of surface passivation on the electronic structure was investigated by coordinating neutral ligands to the 86-atom HgTe cluster. Four representative ligand types were considered: methylamine (CH3NH2), methanethiol ($CH_3SH$), methylphosphine ($CH_3PH_2$), and methanol ($CH_3CH_2OH$). These ligands were selected to probe the influence of different donor atoms (N, S, P, and O) and ligand–surface interactions on the frontier electronic structure. [26–28]

Conformer generation for all studied ligands was performed using the RDKit with the MMFF94s forcefield.[48,49] The lowest-energy conformer for each ligand was subsequently placed in the center of a rectangular simulation box with at least 10 Å of vacuum in each Cartesian direction and relaxed using VASP.

Ligands were coordinated to undercoordinated surface Hg atoms to model passivation of electronically active surface regions. Multiple ligand coverages ranging from three to six ligands per surface region were examined. In addition, adsorption on distinct surface domains (denoted Face 1 and Face 2) was considered in order to evaluate the influence of local coordination geometry and surface polarity on ligand-induced electronic-structure modification.

All ligand-passivated structures were fully relaxed prior to electronic-structure analysis. This approach enables systematic evaluation of how ligand identity, adsorption geometry, and local coordination environment influence suppression of surface-associated edge states and modulation of the HOMO–LUMO gap. The selected ligand set spans a range of donor strengths and orbital interactions, from relatively weak oxygen-based ligands to stronger σ-donor species such as amines, thereby allowing trends in ligand-surface hybridization and frontier-state tuning to be identified.

Because the ligand-passivated structures were fully relaxed, $E_{pass}$ should be interpreted as an overall passivation stabilization energy rather than an isolated single-bond strength. This quantity includes ligand–surface coordination, relaxation of the HgTe surface, ligand-shell rearrangement, and possible cooperative interactions between ligands. The values are therefore most useful for comparing relative stabilization trends among ligand types and adsorption motifs calculated using the same protocol.

## Acknowledgements

R. A. appreciates the support of the Director's Postdoctoral Fellowship at Los Alamos National Laboratory (LANL), which is funded by the Laboratory Directed Research and Development (LDRD) at LANL. This research is supported by the U.S. Department of Energy (DOE), Office of Science, Basic Energy Sciences, as part of the CHIME, a Microelectronics Science Research Center (MSRC). The work at LANL was performed in part at the Center for Integrated Nanotechnologies (CINT), a U.S. DOE and Office of Basic Energy Sciences user facility. This research used resources provided by the LANL Institutional Computing Program. LANL is operated by the US Department of Energy NNSA under Contract No. 89233218CNA000001. P.J.L acknowledges the support of the U.S. DOE through the LANL/LDRD Program and the Center for Nonlinear Studies (CNLS) for this work.